\documentclass[reprint, amsmath,amssymb, aps]{revtex4-2}

\usepackage{graphicx}
\usepackage{dcolumn}
\usepackage{bm}
\usepackage[left]{lineno}

\begin{document}
\title{Charge transport in a polar metal}
\author{Jialu Wang$^1$, Liangwei Yang$^1$, Carl Willem Rischau$^2$, Zhuokai Xu$^1$, Zhi Ren$^1$, Thomas Lorenz$^3$, Joachim Hemberger$^3$} \author{Xiao Lin$^{1}$}\email{Correspondence: linxiao@westlake.edu.cn}

\author{Kamran Behnia$^{3,4}$}

\affiliation{$^{1}$ School of Science, Westlake Institute for Advanced Study, Westlake University, 18 Shilongshan Road, 310024 Hangzhou, China\\
$^{2}$ Department of Quantum Matter Physics, University of Geneva, 1205 Geneva, Switzerland\\
$^{3}$ II. Physikalisches Institut, Universit\"{a}t zu K\"{o}ln, Z\"{u}lpicher Str. 77, 50937 K\"{o}ln, Germany \\
$^{4}$ Laboratoire Physique et Etude de Mat\'{e}riaux (CNRS-UPMC), ESPCI Paris, PSL Research University, 75005 Paris, France}

\date{\today}

\begin{abstract}
The fate of electric dipoles inside a Fermi sea is an old issue, yet poorly-explored. Sr$_{1-x}$Ca$_x$TiO$_{3}$ hosts a robust but dilute ferroelectricity in a narrow ($0.002<x<0.02$) window of substitution. This insulator becomes metallic by removal of a tiny fraction of its oxygen atoms. Here, we present a detailed study of low-temperature charge transport in  Sr$_{1-x}$Ca$_x$TiO$_{3-\delta}$, documenting the evolution of resistivity with increasing carrier concentration ($n$). Below a threshold carrier concentration, $n^*(x)$, the polar structural phase transition has a clear signature in resistivity and Ca substitution significantly reduces the 2 K mobility at a given carrier density. For three different Ca concentrations, we find that the phase transition fades away when one mobile electron is introduced for about $7.9\pm0.6$ dipoles. This threshold corresponds to the expected peak in anti-ferroelectric coupling mediated by a diplolar counterpart of Ruderman-Kittel-Kasuya-Yosida (RKKY) interaction. Our results imply that  the transition is driven by dipole-dipole interaction, even in presence of a dilute Fermi sea. Charge transport for $n < n^*(x)$ shows a non-monotonic temperature dependence, most probably caused by scattering off the transverse optical phonon mode. A quantitative explanation of charge transport in this polar metal remains a challenge to theory.  For $n\geq  n^*(x)$,   resistivity follows a T-square behavior together with slight upturns (in both Ca-free and Ca-substituted samples).  The latter are reminiscent of Kondo effect and most probably due to oxygen vacancies.

\end{abstract}

\maketitle

\section{Introduction}

The concept of a polar or 'ferroelectric' metal was first proposed by Anderson and Blount in 1960s \cite{Anderson1965}. They considered a continuous structural phase transition breaking the inversion symmetry and leading to the appearance of a polar axis in a metal. This appears counter-intuitive since one expects mobile electrons to strongly screen electric field. Recently, however,  'ferroelectric' metallicity was reported in  LiOsO$_3$ \cite{Shi2013}.  It was found that this stoichiometric metal shows a structural phase transition to non-centrosymmetric  rhombohedral phase (R3c) below 140 K. The polar structural transition, which manifests itself as a kink in resistivity of this metal, is analogue to what occurs in its insulating ferroelectric cousins LiNbO$_3$ and LiTaO$_3$ \cite{Shi2013}. Puggioni and Rondinelli  \cite{Puggioni2014} have argued that such polar metals can be found when there is an unusually weak coupling between mobile electrons and transverse optical phonons, which drive ferroelectricity.  This scenario has been documented in LiOsO$_3$ by recent ultrafast spectroscopy measurements \cite{Laurita2019}.

Paraelectric solids close to a ferroelectric transition \cite{Kvyatkovskii2001} provide an alternative platform for  a meeting between metallicity and ferroelectricity. One example is PbTe, a narrow-gap semiconductor close to a ferroelectric instability. The unavoidable presence of doping defects makes available PbTe samples dilute metals. Isovalent substitution of Pb by Ge leads to a structural phase transition from cubic to a non-centrosymmetric rhombohedral phase \cite{Hohnke1972}, which would have been ferroelectric in absence of mobile electrons. The ferroelectric-like transition was revealed by X-ray diffraction, inelastic neutron and Raman scattering \cite{Hohnke1972,Alperin1972,Sugai1979}. Charge transport in presence of local dipoles was studied several decades ago \cite{Takaoka1979,Yaraneri1981,Takano1984,Katayama1987}.

SrTiO$_3$ single crystals, in contrast to PbTe, can be made stoichiometric enough to be insulating. Proximity to a ferroelectric quantum critical point \cite{Narayan2019,Chandra2017} is manifested by a large electric permittivity of SrTiO$_3$ ($\varepsilon_\textrm{r}>20000$) \cite{Muller1979}. As a consequence, this insulator displays a number of intriguing properties \cite{Martelli}. A FE state emerges upon substitution of a tiny fraction of Sr ions with Ca \cite{Bednorz1984}. Moreover, this quantum paraelectric can become a  dilute metal  (with a carrier concentration as small as $\approx10^{16}$ cm$^{-3}$) upon oxygen reduction \cite{Spinelli2010,Bhattacharya2016}. The dilute metal undergoes a superconducting transition below 0.3 K \cite{Koonce1967,Lin2013, Collignon2018}.

Rischau \textit{et al.} have recently found that a superconducting phase coexists with a FE-like instability in n-doped Sr$_{1-x}$Ca$_x$TiO$_{3-\delta}$ and superconductivity and ferroelectricity (FE)  may be intimately linked \cite{Willem2017}. The FE transition of  insulating  Sr$_{1-x}$Ca$_x$TiO$_{3}$ was found to survive in metallic Sr$_{1-x}$Ca$_x$TiO$_{3-\delta}$. Being a metal, the latter does not show a bulk reversible electric polarization and cannot be a true ferroelectric. Nevertheless, it shows anomalies in various physical properties at the Curie temperature of the insulator. For example, Raman scattering found that the hardening of the FE soft mode in the dilute metal is indistinguishably similar to what is seen in the insulator \cite{Willem2017}. The anomaly in resistivity was found to terminate at a threshold carrier density ($n^*$), near which the superconducting transition temperature was enhanced \cite{Willem2017} providing evidence for a link between superconducting pairing and ferroelectricity, a subject of present attention \cite{Stucky2016,Rowley2018,Herrera2018,Tomiokal2019,Ahadi2019,Edge2015,Kedem2016,Wolfle2018,Kanasugi2019,Marel2019,Kozii2019}.

In this paper, we present a study of low-temperature electrical resistivity in dozens of Sr$_{1-x}$Ca$_x$TiO$_{3-\delta}$ single-crystals with $x=0$, $0.22\%$, $0.45\%$, $0.9\%$ documenting in detail the evolution of charge transport with increasing concentrations of electric dipoles and charge carriers, a question, which was not addressed in depth by a previous study \cite{Willem2017}. In Sr$_{1-x}$Ca$_x$TiO$_{3-\delta}$  metallicity and ferroelectricity are both dilute. Therefore, the distance between dipoles and mobile electrons can be separately tuned but kept  much longer than the interatomic distance. We  find that the magnitude of low-temperature mobility is significantly reduced below $n^*(x)$ and the mean-free-path gently peaks near $n^*(x)$, where the quadratic temperature dependence of resistivity is restored. Moreover, we find that $n^*$ is proportional to $x$, implying the threshold density occurs at a fixed ratio between the inter-carrier and the inter-dipole distance. We will argue that these features are all consistent with the hypothesis of a dipolar RKKY interaction, which was theoretically proposed a quarter-century ago \cite{Glinchuk1992}. At much higher carrier densities  ($n>5\times10^{19}$ cm$^{-3}$), we observe a slight upturn in low-temperature resistivity of Ca-free and Ca-substituted samples and attribute it to a Kondo effect associated with oxygen vacancies.

\section{Results}

\begin{figure}
\includegraphics[width=9cm]{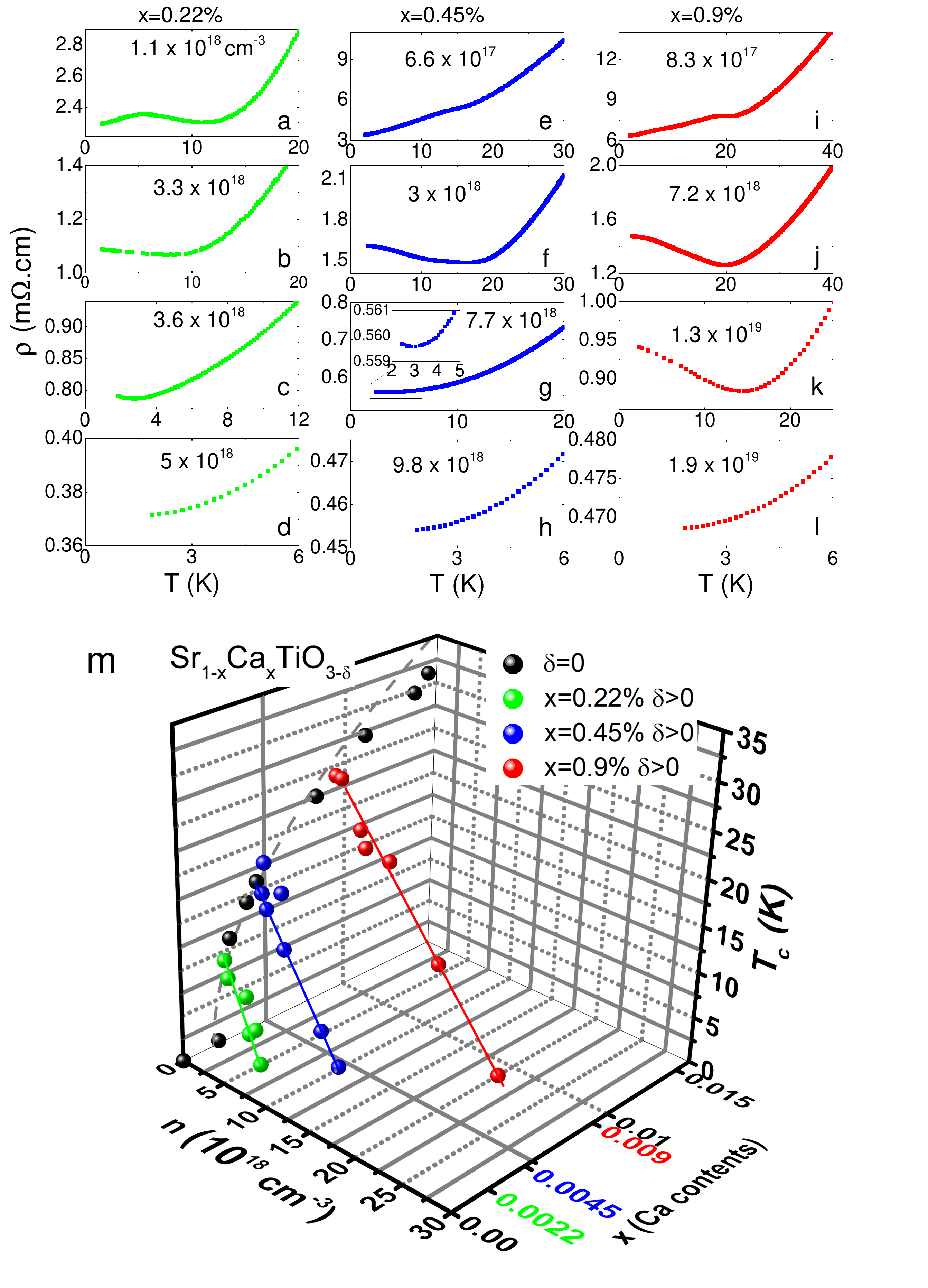}
\caption{The evolution of polar structure phase transition as the function of Ca contents and electron doping. \textbf{Upper panel}: Temperature dependence of resistivity for Sr$_{1-x}$Ca$_x$TiO$_{3-\delta}$. \textbf{a-d}: $x=0.22\%$; \textbf{e-h}: $x=0.45\%$; \textbf{i-l}: $x=0.9\%$. \textbf{Lower panel(m)}: $T_\textrm{c}$ of FE transition or polar structural phase transition as the function of $x$ and $n$. The black balls mark $T_\textrm{c}$  of insulating Sr$_{1-x}$Ca$_x$TiO$_3$ samples \cite{Bednorz1984}. The green, blue, red balls mark the $T_\textrm{c}$ of polar structural phase transitions in n-doped Sr$_{1-x}$Ca$_x$TiO$_{3-\delta}$ with $x=0.22\%$, $0.45\%$ and $0.9\%$ respectively.}
\end{figure}

The upper panel (a-l) of Fig. 1 plots the low temperature dependence of resistivity for Sr$_{1-x}$Ca$_x$TiO$_{3-\delta}$ at low $n$ and $x=0.22\%$, $0.45\%$, $0.9\%$. In Fig. 1a, 1e and 1i, resistivity shows an anomaly at lowest n for all three groups of samples. Increasing n-doping, the anomaly shifts to lower temperatures, evolves into an minimum and finally disappears at a threshold doping ($n^*(x)$), seen in Fig. 1b-1d, 1f-1h and 1j-1l. Rischau \textit{et al.}  found that in selected samples at $x=0.2\%$ and $0.9\%$, this anomaly occurs close to where the structural phase transition was detected by Raman spectroscopy, sound velocity and thermal expansion measurements  \cite{Willem2017}. Most recently, thermal expansion measurements documented the evolution of the structural phase transition in Sr$_{1-x}$Ca$_x$TiO$_{3-\delta}$ ($x=0.9\%$), starting from the insulating phase and extending deep into the metallic phase \cite{Engelmayer2019}. The study found that below $n^*$, the transition temperature and the magnitude of the transition-induced anomaly continuously decreased with increasing carrier concentration. Above $n^*$, a small residual anomaly with a concentration-independent temperature scale was observed to survive. No sign change in the thermal expansion coefficient ($\alpha$) was observed at $n^*$. Thus, the corresponding Gr\"uneisen ratio $\alpha/C_p$ ($C_p$ is the specific heat) does not change sign. This implies either the absence of a quantum critical point at $n^*$ or its insensitivity to uniaxial pressure\cite{Engelmayer2019}.

Fig. 1m shows a 3D plot of the phase diagram of Sr$_{1-x}$Ca$_x$TiO$_{3-\delta}$ with two tunable parameters ($n$ and $x$). In the $n=0$ plane, Sr$_{1-x}$Ca$_x$TiO$_3$ becomes ferroelectric when $x_\textrm{c}>0.18\%$. Above this critical concentration of Ca substituents, $T_\textrm{c}$ scales with $x$ following $T_\textrm{c}\thicksim|x-x_\textrm{c}|^{1/2}$ \cite{Bednorz1984}. The anomaly in resistivity of n-doped samples is shown with green, blue and red symbols for $x=0.22\%$, $0.45\%$, $0.9\%$ respectively. The anomaly caused by the structural phase transition (strictly speaking, only a true FE state at $n=0$) shifts to lower temperatures with increasing $n$. At a threshold doping  ($n^*(0.22\%)\approx5\times10^{18}$ cm$^{-3}$, $n^*(0.45\%)\approx1\times10^{19}$ cm$^{-3}$ and $n^*(0.9\%)\approx2\times10^{19}$ cm$^{-3}$),  resistivity becomes metallic down to lowest temperatures (see panels, 1d, 1h, 1l).

\begin{figure}
\includegraphics[width=8cm]{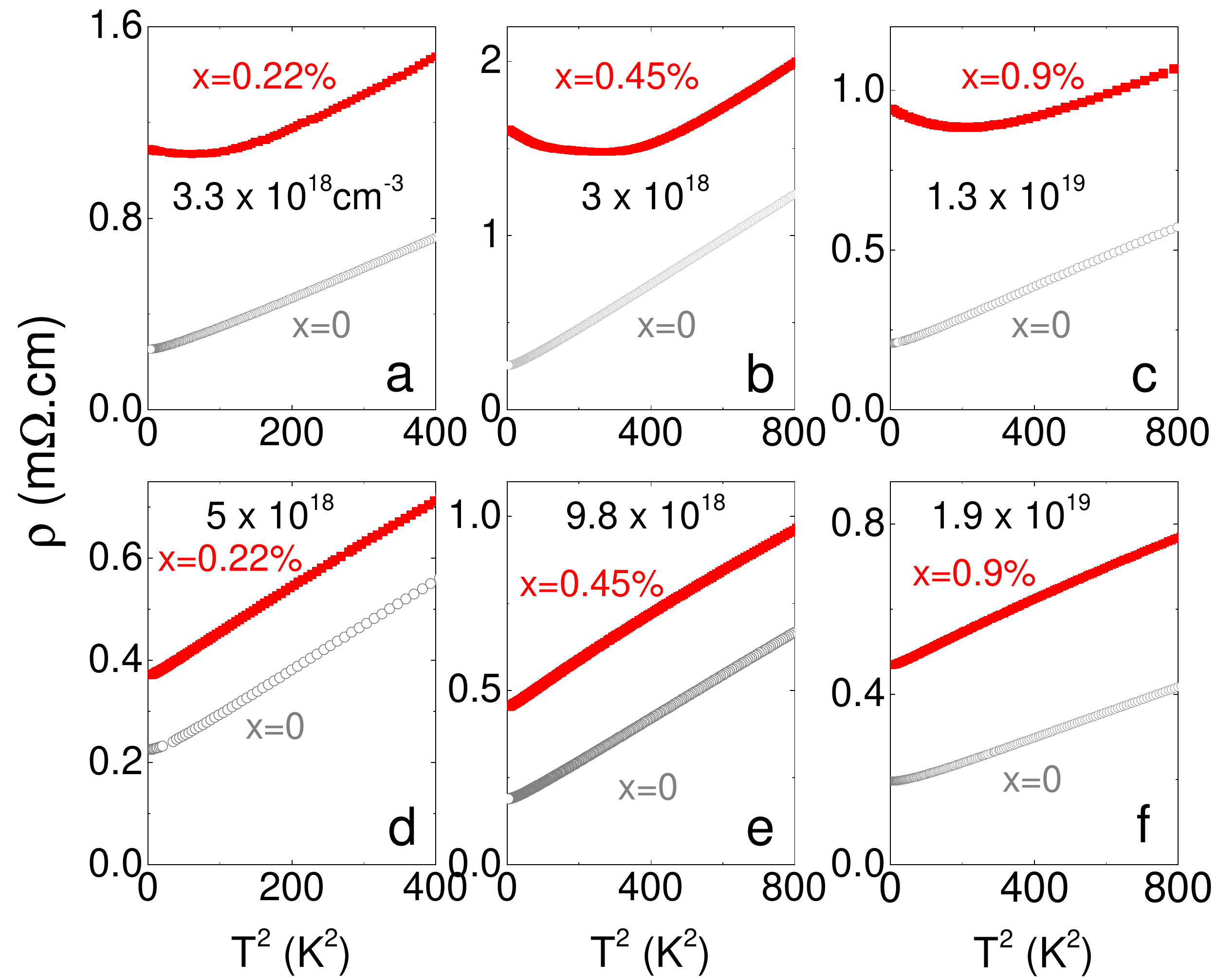}
\caption{Resistivity plotted as a function of $T^2$ at low temperatures for Sr$_{1-x}$Ca$_x$TiO$_{3-\delta}$ with $x=0.22\%$, $0.45\%$ and $0.9\%$, compared with $x=0$ at similar carrier densities. When $n<n^*(x)$ (\textbf{a-c}) , the Ca-substituted samples show a strong deviation from the $T^2$ temperature dependence seen in Ca-free samples.  On the other hand, when $n\approx n^*(x)$ (\textbf{d-f}), the $T^2$ temperature dependence is maintained in Ca-substituted samples and the slope remains the same.}
\end{figure}

The low-temperature resistivity of strontium titanate follows a simple quadratic temperature dependence: $\rho=\rho_0+AT^2$ \cite{Lin2015,Marel2011}. Elastic scattering is represented by residual resistivity, $\rho_0$,  intimately linked to the asymptotic low-temperature mobility set by disorder. Inelastic scattering among electrons is expected to give rise to the  $AT^2$ term. However, the quadratic temperature dependence of resistivity has several puzzling features. One is that it persists even in the dilute regime where the Fermi surface is too small to allow Umklapp scattering among electrons \cite{Lin2015}. The other is that this T-square resistivity extends beyond the degeneracy temperature and the exponent gradually enhances with warming \cite{Lin2017}. Nevertheless, the magnitude of T-square resistivity prefactor corresponds to what is found empirically in other metals (dilute or correlated) with a similar Fermi energy \cite{Collignon2018}. Moreover, this prefactor evolves smoothly and universally with carrier concentration and has a similar amplitude in other quantum paralectric perovskite oxides \cite{Engelmayer2019b} such as EuTiO$_3$ and KTaO$_3$ in spite of their different phonon spectra. The origin of this T-square resistivity is  still debated \cite{Maslov2017, Collignon2018}.

Both the elastic and inelastic terms of resistivity are affected by the passage across $n^*(x)$. The quadratic temperature dependence is drastically altered upon the introduction of Ca atoms.  In Fig. 2, the resistivity of Sr$_{1-x}$Ca$_x$TiO$_{3-\delta}$ with $x=0.22\%$, $0.45\%$ and $0.9\%$, is plotted \textit{vs.} $T^2$ and compared with Ca-free samples of similar $n$. As seen in the upper panel, which shows typical data for $n<n^*(x)$, the resistivity shows a minimum followed by an upturn, in presence of the structural phase transition. Note that the non-monotonic temperature dependence of mobility cannot be a simple consequence of the temperature dependence of the electric permittivity (See Supplementary Figure 1). In contrast, in SrTiO$_{3-\delta}$ with similar $n$, $T^2$ dependence of resistivity persists down to lowest temperatures \cite{Lin2015,Marel2011}.  The lower panel of Fig. 2 shows the behavior at $n \simeq n^*(x)$. One can see that the $T^2$ resistivity of Ca-doped samples is restored. For more data at $n>n^*(x)$, please refer to  Supplementary Figure 2. Moreover, the slope  is similar in Ca-substituted and Ca-free samples. In other words, the prefactor of quadratic resistivity, which strongly depends on carrier concentration but not on residual resistivity, is similar in Ca-substituted and Ca-free samples at the same carrier density. Let us note that this strict Fermi-liquid behavior at $n^*(x)$, indicates that is not a 'quantum critical point' where a 'non-Fermi-liquid behavior' is commonly sought and often found. This is in agreement with the absence of a sign change in thermal expansion\cite{Engelmayer2019}. The other implication of this observation is that below $n^*(x)$,  electrons suffer additional inelastic scattering in presence of aligned dipoles. The destruction of the structural phase transition at $n^*(x)$ suppresses this additional mechanism and restores the $T^2$ resistivity associated with electron-electron scattering \cite{Lin2015}. This implies the existence of additional scattering  in the polar metallic state, totally smearing T-square resistivity.

\begin{figure*}
\centering
\includegraphics[width=14cm]{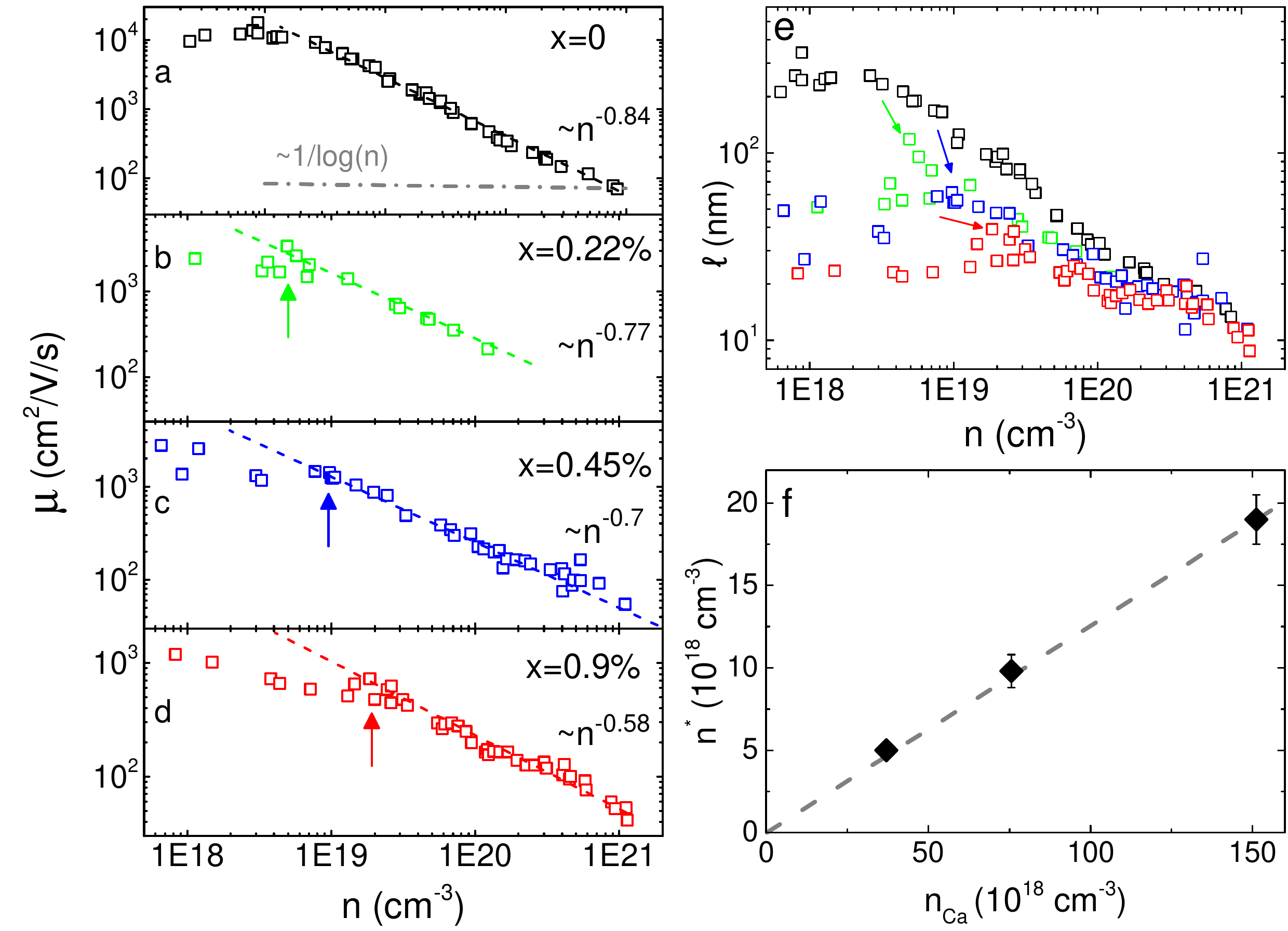}
\caption{The Hall mobility and mean-free-path for three groups of Sr$_{1-x}$Ca$_x$TiO$_{3-\delta}$ with electron doping. \textbf{a-d}: The Hall mobility ($\mu$) at 2 K as a function of $n$ in Log-Log scale for Sr$_{1-x}$Ca$_x$TiO$_{3-\delta}$ with $x=0$, $0.22\%$, $0.45\%$ and $0.9\%$. The dashed lines are fits to power laws. Arrows mark the threshold doping ($n^*(x)$) for Ca-doped samples. The dash-dot-line in \textbf{a} represents $\frac{1}{\textrm{log}(n)}$. \textbf{e}: Mean-free-path ($\ell$) at 2 K as a function of $n$ in Log-Log scale for four sets of samples. The gentle maxima are marked by arrows at a carrier density corresponding to $n^*(x)$. \textbf{f}: $n^*$ as the function of Ca concentration $n_{\textrm{Ca}}=\frac{x}{a_0^3}$, where $a_0=0.3905$ nm is the lattice constant of SrTiO$_3$. The dashed line is a linear fit. The error bars mark the extent of uncertainty in extracting $n^*$ from the fit in Fig. 1m and Fig. 3b-3d.}
\end{figure*}

The presence of FE-like order affects the elastic scattering of the carriers too. The Hall mobility ($\mu$) at 2K for Sr$_{1-x}$Ca$_x$TiO$_{3-\delta}$ with $x=0$, $0.22\%$, $0.45\%$, $0.9\%$ is presented in Fig. 3. The mobility $\mu$ is extracted from Hall resistivity $\rho_{\textrm{yx}}$ and longitudinal resistivity $\rho_{\textrm{xx}}$ using $\mu=\frac{\rho_{\textrm{yx}}}{\rho_{\textrm{xx}} B}$. As seen in the figure, $\mu$ increases with the decreasing in the carrier density ($n$) following an approximate power law $\mu \sim n^{-\alpha}$ shown by the dashed lines. For SrTiO$_{3-\delta}$, the power law behaviour persists down to $n\thicksim2\times10^{18}$ cm$^{-3}$. Below this concentration, it  begins to saturate and then drops at even lower carrier concentrations with the approach of the metal-insulator transition\cite{Lin2013,Spinelli2010}.

In many doped semiconductors, the mobility decreases with increasing carrier concentration \cite{Ellmer2001,Tahar1998}. This has been often discussed in the framework of ionized impurity scattering \cite{Conwell1950}. Several features distinguish the 2K mobility of metallic strontium titanate from ordinary doped semiconductors. First of all, the dependence of mobility with carrier concentration is very steep (i.e. $\alpha$ in $\mu \sim n^{-\alpha}$ is close to unity). Second, this mobility is strongly temperature dependent and passes from a room-temperature value of 5 cm$^2$V$^{-1}$s$^{-1}$ to 20000 cm$^2$V$^{-1}$s$^{-1}$ at liquid He temperature\cite{Tufte1967,Lin2017}. This latter value implies that the low-temperature carrier mean-free-path ($\ell$) becomes much longer than the interdopant  distance ($l_{\textrm{ee}}=n^{-1/3}$).

Dingle provided an expression for low-temperature mobility in a degenerate doped semiconductor, based on the Born approximation for scattering\cite{Dingle1955}. The formula can be written  as $\mu \propto \frac{\textrm{e}}{\hbar} \frac{a_\textrm{B}^*{}^2}{\textrm{log}(n a_\textrm{B}^*{}^3)}$, where $n$ is carrier concentration and $a_\textrm{B}^*$ is the effective Bohr radius. This expression does not seem to provide a good description of our data, as seen in Fig. 3a.

Recently, it has been argued \cite{Behnia2015} that these peculiar features of mobility in dilute metallic strontium titanate can be traced back to the long effective Bohr radius, $a_\textrm{B}^*$, of the parent insulator. The large electric permittivity \cite{Muller1979} elongates the Bohr radius to 600 nm, which is to be compared to 1.5 nm in silicon. This in turn affects the Thomas-Fermi  screening length of the metal, which depends on it:
\begin{equation}
r_{\textrm{TF}}=\sqrt{\frac{\pi a_\textrm{B}^*}{4k_\textrm{F}}}
\end{equation}

The combination of a large $a_\textrm{B}^*$ and a small $k_\textrm{F}$ elongates the screening length and therefore short-distant irregularities in the dopant distribution are smoothed out. This simple approach yields this expression for mobility \cite{Behnia2015}: 
\begin{equation}
\mu \propto a_\textrm{B}^*{}^{1/2}n^{-5/6}
\end{equation}

Fig. 3a confirms that this expression gives a surprisingly good account of the variation of 2 K mobility with $n$ in  SrTiO$_{3-\delta}$ \cite{Behnia2015}.  As seen in Fig. 3b-3d, Ca substitution leads to a slight decrease in the exponent of the power law exponent ($\alpha$). More importantly, a clear deviation from the power-law behaviour occurs below $n^*(x)$ marked by vertical arrows. In other words,  the carrier mobility is significantly reduced when the system orders.  Fig. 3e shows the mean-free-path ($\ell$) calculated through $\ell=\mu\hbar k_\textrm{F}/\textrm{e}$, in which $\mu$ is taken from the Hall mobility, $k_\textrm{F}$ is the Fermi wave vector deduced from $k_\textrm{F}=(3\pi^2 n)^{1/3}$ by assuming a single isotropic Fermi surface. It presents a mild maximum marked by three arrows at $n^*(x)$. Well above  this threshold density of $n^*(x)$, the mean-free-path of the Ca-doped and Ca-free samples gradually merges. Thus, we conclude that the presence of the FE-like order inside the metal has drastic consequences for elastic scattering of electrons too.

Fig. 3f shows $n^*$ extracted from our data as a function of $n_{\textrm{Ca}}$. One can see that the two concentrations are proportional to each other. In other words, the threshold inter-electron distance ($l^*_{\textrm{ee}}$) linearly scales with the average distance between Ca ions ($n_{\textrm{Ca}}^{-1/3}$), the slope of which indicates that the destruction of the FE-like order happens when there is one mobile electron per 7.9$\pm$0.6 Ca ions.

\section{DISCUSSION}

Thus, we find that below a threshold concentration: i) an additional mechanism for inelastic scattering sets in; ii) the low-temperature mobility  is significantly reduced. Moreover: iii) this  threshold density  for the destruction of the polar metal is proportional to Ca concentration and  iv) at this density carrier mean-free-path gently peaks. We are now going to argue that these observations support a picture in which off-center Ca sites generate electric dipoles interacting with each other inside a Fermi sea.

\begin{figure}
\includegraphics[width=6cm]{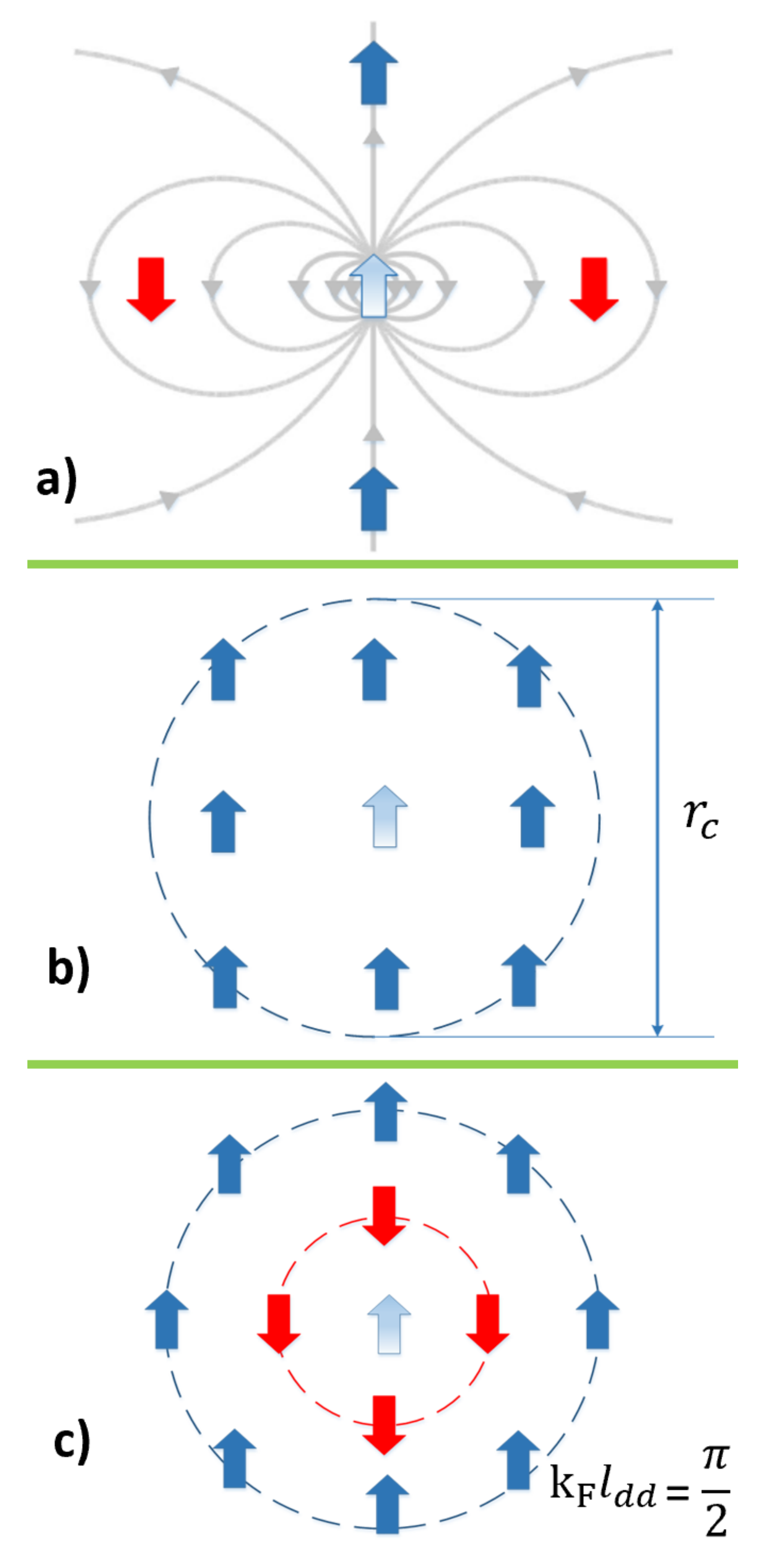}
\caption{Dipole-dipole interaction in three contexts. The central dipole (open arrow) can interact either ferroelectrically (in blue) or anti-ferroelectrically with dipoles (in red). \textbf{a}: In an ordinary ionic crystal, the interaction is ferroelectric along one orientation and anti-ferroelectric along the perpendicular one, impeding the emergence of long-range ferroelectricity. \textbf{b}: In a highly-polarizable lattice, the interaction is ferroelectric over a length scale, r$_\textrm{c}$, longer than the inter-dopant distance, $l_{\textrm{dd}}$. This allows the formation of large ferroelectric cluster composed of many dipoles. The typical size of these clusters is set by r$_c$ \cite{Vugmeister1980,Vugmeister1990,Samara2003}. \textbf{c}: In the presence of a Fermi sea, ferroelectric and anti-ferroelectric coupling between dipoles alternate radially. According to our observations, anti-ferroelectric coupling between dipoles destroys the ferroelectric order in  Sr$_{1-x}$Ca$_x$TiO$_{3-\delta}$ at a carrier density of $n^*(x)$, which corresponds to $2k_\textrm{F} l_{\textrm{dd}}=\pi$.}
\end{figure}

Let us begin with a fundamental question: what drives the FE transition in the insulating  Sr$_{1-x}$Ca$_x$TiO$_3$ for $0.0018<x<0.02$? According to the scenario invoked by previous authors \cite{Bednorz1984,Kleemann2000}, an off-site Ca atom generates an electric dipole, which couples to the soft transverse optical phonon of the host lattice. Theory \cite{Vugmeister1980} had predicted that cooperative phenomena between paraelectric defects in an easily polarizable crystal lead to polar clusters whose size is set by the polar correlation length of the host lattice ($r_\textrm{c}$). The latter depends on the velocity and the frequency of the soft mode of the host polarizable lattice ($r_\textrm{c}=v_\textrm{s}/\omega_\textrm{0}$) \cite{Vugmeister1990,Samara2003}. With cooling $r_\textrm{c}$ increases, the interaction between clusters becomes stronger and at a sufficiently low temperature, cluster percolation leads to a ferroelectric order. A transverse Ising model, treating dipoles as pseudospins, has been successful in describing the Sr$_{1-x}$Ca$_x$TiO$_3$ phase diagram \cite{Hemberger1996,Wang1998}.

In an ordinary ionic crystal, there is no mean-field basis for favoring ferroelectricity. This is because the electric field generated by a dipole does favor opposite alignments for a neighboring dipole along perpendicular orientations (Fig. 4a). In an easily polarizable crystal, on the other hand,  the interaction becomes ferroelectric over a polarization length ($r_\textrm{c}$), thanks to the presence of soft transverse optic phonon modes. Since $r_\textrm{c}$ is significantly longer than the typical distance between dipolar impurities, sizable ferroelectric clusters \cite{Vugmeister1980,Vugmeister1990,Samara2003} (Fig. 4b) become energetically stable. This paves the way for the emergence of long-range order.

This approach provides a satisfactory explanation for two experimental observations. The first is the presence of a ferroelectric order triggered by off-center impurities in quantum paralectrics such as SrTiO$_3$ and KTaO$_3$ \cite{Schremmer} and its absence in ordinary ionic crystals with no soft mode such as alkali halide crystals \cite{Vugmeister1990}. The second observation is the existence of a critical concentration for the emergence of ferroelectricity  in Ca-substituted SrTiO$_3$ \cite{Bednorz1984}. Let us note, however, that extrinsic Ca ions in strontium titanate lattice can also interact with each other via the strain field, they create by reducing the size of the lattice cell which they occupy. The latter feature has been invoked to explain the weakening of FE order for $x<0.02$ \cite{Zhang2002}.

In this context, our observations raise two questions specific to metallic Sr$_{1-x}$Ca$_x$TiO$_{3-\delta}$. First of all, at small $n$, how can a phase transition driven by the dipolar interaction among Ca sites hardly be affected  by the capacity of mobile electrons supposed to screen any electric field?  A second question arises by the existence of $n^*(x)$. Why does the order  eventually fade away and what does this threshold density correspond to?

The first question brings to mind a remark by Landauer \cite{Landauer} according to which, neither electric field nor carrier concentration can be strictly homogeneous in a metal containing defects. This is particularly true in our case, where the Thomas-Fermi screening length is very long by a combination of long Bohr radius and small Fermi wave-vector. When the carrier density is as low as $10^{18}$ cm$^{-3}$,  Eq. 1 yields $r_{\textrm{TF}}\simeq$80 nm,  much longer than the distance between the dipoles or the $r_\textrm{c}$ of the insulator. The Fermi sea is too dilute to impede dipolar interaction. The attenuation of mobility implies enhanced electric-field inhomogeneity brought by the alignment of dipoles, providing additional support for picturing persistent electric interaction in spite of a Fermi sea.

The second question leads us to an additional interaction mechanism offered to electric dipoles in presence of a Fermi sea (see Fig. 4c). This was proposed first by Glinchuk and Kondakova in 1992 \cite{Glinchuk1992}. The opposite charges of an electric dipole are expected each to generate Friedel oscillations \cite{Friedel} inside a Fermi sea, which is a distinct source of dipolar interaction. Indeed, it is established  that  localized magnetic spins inside a Fermi sea generate similar oscillations and interact through what is known as the Ruderman-Kittel-Kasuya-Yosida (RKKY) mechanism \cite{Ruderman1954}. Glinchuk and Kondakova proposed  a dipolar analogue of RKKY interaction. This dipole-dipole interaction (V$_{\textrm{dd\_R}}$) depends on the magnitude of the electric dipole moment and has an alternating sign as a function of the inter-dipole distance ($l_{\textrm{dd}}$) and Fermi wave vector ($k_\textrm{F}$) \cite{Glinchuk1992}:

\begin{equation}
V_{\textrm{dd\_R}}\propto \frac{\textrm{cos}(2k_\textrm{F}l_{\textrm{dd}})}{l_{\textrm{dd}}^3}
\end{equation}

This expression for interaction between two dipoles is to be compared with what is expected in vacuum:
\begin{equation}
V_{\textrm{dd\_V}}\propto\frac{1}{l_{\textrm{dd}}^3}
\end{equation}
or in a highly-polarizable insulator with a polarization correlation radius of $r_\textrm{c}$  \cite{Vugmeister1990}:
\begin{equation}
V_{\textrm{dd\_C}}\propto\frac{exp(-l_{\textrm{dd}}/r_\textrm{c})}{l_\textrm{dd}r_{\textrm{c}}^2}
\end{equation}

According to Eq. 3, the parallel alignment of nearest dipoles becomes energetically unfavorable when $\textrm{cos}(2k_\textrm{F}l_{\textrm{dd}})=-1$ or $2k_\textrm{F}l_{\textrm{dd}}= \pi$. Assuming an isotropic Fermi surface ($k_\textrm{F}=(3\pi^2n)^{1/3}$), one finds that this will happen when  $\frac{n_\textrm{d}}{n}=\frac{24}{\pi}$. In other words, in presence of a single-band isotropic metal, the antiferroelectric coupling between neighboring dipoles is expected to peak when there is $\frac{24}{\pi}\simeq7.6$ dipoles per mobile electrons.

Thus, the simplest picture of RKKY-like interaction between electric dipoles expects a proportionality between dipole density and the charge carrier density required to achieve maximum anti-ferroelectric coupling between neighboring dipoles and this proportionality is simply close to 8 (which is because we are in three dimensions and for each dimension one needs two dipoles per one electron for destructive interference). It is worth noting the following two facts. First of all: Eq. 3  does not include the angular dependence of the interaction between neighboring dipoles, which can have different orientations. Second: the dopants are randomly distributed in the real system, which should be taken account in a more sophisticated theoretical model. Thus, one may conjecture that the $\frac{n_\textrm{d}}{n}$ ratio remains the same after averaging the space and angle dependence of interactions. If not, it would be an amazing coincidence that for three different Ca contents, the experimentally observed $\frac{n_\textrm{d}}{n}$ corresponds to what is expected in the simple approach neglecting these details.  Further theoretical efforts are required to clarify this issue.

One may be tempted by an alternative picture in which the FE-like transition is destroyed because the screening length shrinks with increasing carrier concentration. However, since this evolution is very slow ($r_{\textrm{TF}}\propto n^{-1/6}$), it would be very hard to explain the linear proportionality between $n^*$ and $x$.

As mentioned above, when carrier density is below $n^*(x)$, we are in presence of a polar metal where mobile electrons and aligned dipoles co-exist. Explaining the transport properties of the metal in this regime emerges as a  challenge to future theoretical works. The complex temperature dependence of electrical resitivity (markedly different from quadratic) is to be understood in a picture where electrons are scattered off a hardened ferrolectric mode.

\begin{figure}
\includegraphics[width=8cm]{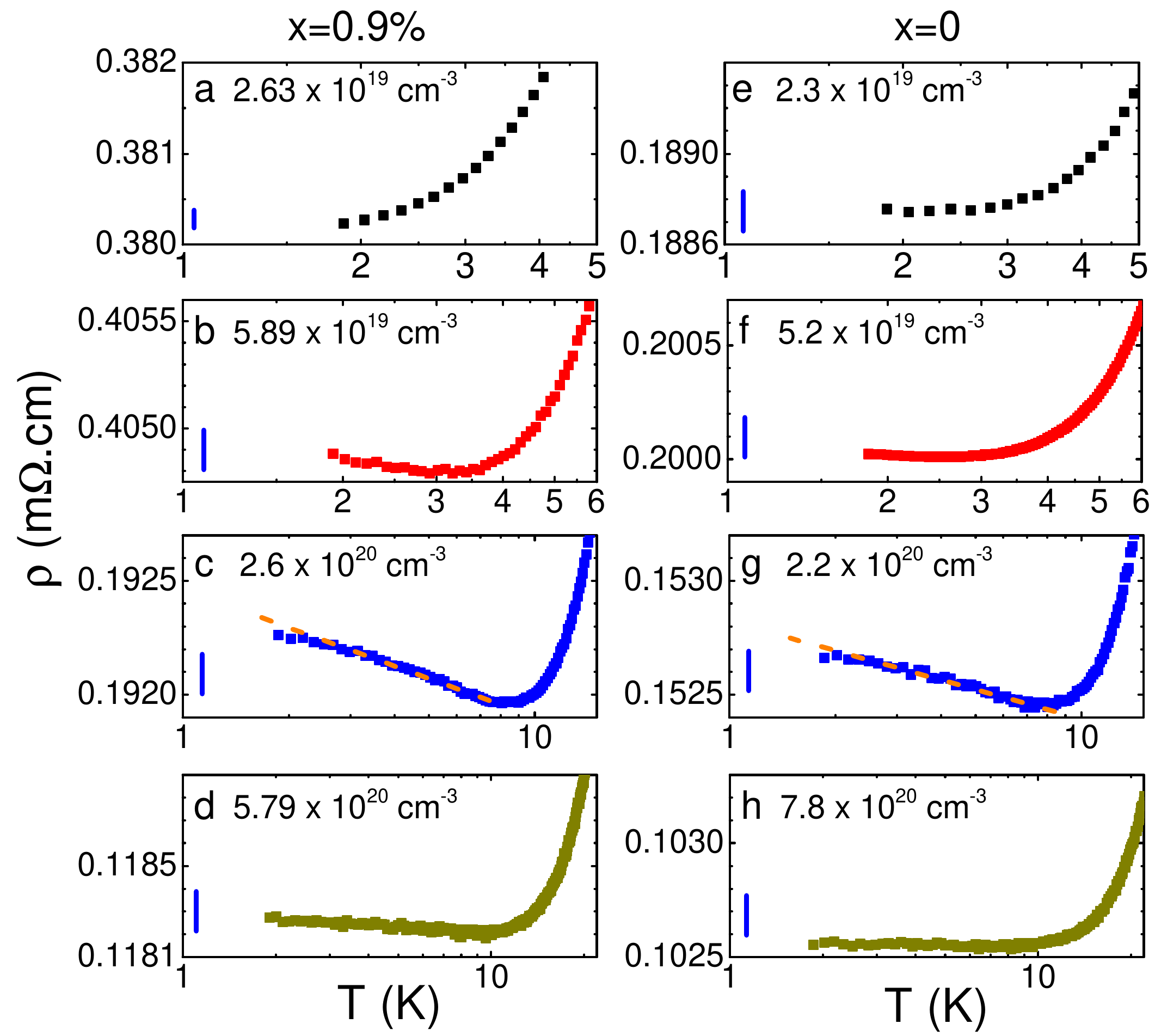}
\caption{Temperature dependence of resistivity at low temperatures  above $n^*$. Data for Sr$_{1-x}$Ca$_x$TiO$_{3-\delta}$ with $x=0.9\%$ (\textbf{a-d}) is compared with $x=0$ (\textbf{e-h}). The vertical blue bars correspond to a value of 0.2 \rm $\mu\Omega$.cm. Note the presence of small upturns in both cases corresponds to a sub-percent change in resistivity.}
\end{figure}

What happens to the electric dipoles when the carrier density exceeds $n^*(x)$? According to the thermal expansion measurements \cite{Engelmayer2019}, a smeared anomaly continues to survive. One may speculate that a non-percolative dipole glass \cite{Vugmeister1990}  persists at higher densities. In our study of resistivity, we cannot see any significant difference between Ca-free and Ca-substituted strontium titanate when  $n> n^*(x)$. On the other hand, we do see small upturns in resistivity, reminiscent of the Kondo effect, which constitutes yet another source of information.

Low-temperature resistivity of Sr$_{1-x}$Ca$_x$TiO$_{3-\delta}$ (with $x=0.9\%$)  above $n^*(x)$ is shown in Fig. 5a-5d. One can see that a less pronounced upturn re-appears above $n^*(x=0.9\%)$. Similar data for Ca-free samples are shown in Fig. 5e-5h for comparison. No clear upturn can be seen in Fig. 5e and 5f with $n$ below $5.2\times10^{19}$ cm$^{-3}$. With further increase of carrier concentration, a small upturn appears. No  difference between Ca-doped and Ca-free samples can be detected. We note however, that these upturns in resistivity correspond to a relative change of several $10^{-3}$. This is orders of magnitudes smaller than what can be seen in Fig. 2 for $n<n^*$. Their small magnitude as well as their presence in Ca-free samples point to an origin which is not the one seen in the presence of the phase transition and discussed above.

As seen in Fig. 5c and 5g, the  temperature dependence of these upturns is roughly logarithmic. The logarithmic temperature dependence of resistivity is reminiscent of the Kondo effect arising when a localized spin couples to a Fermi sea \cite{Kondo1964}. It is now known that the Kondo effect can be present whenever a Two-Level-System (TLS) is embedded in a Fermi sea \cite{Glazman} and experiments \cite{Fernandez,Jarillo,Matsushita} have documented a variety of non-magnetic counterparts of the original Kondo effect. Our upturns in resistivity, both by their magnitude and temperature dependence are reminiscent of what Matsushita \textit{et al.} observed in Tl-doped PbTe and attributed to charge Kondo effect. The valence degeneracy of Tl dopants was identified as the driver of the Kondo effect \cite{Matsushita}.

If this is a Kondo effect, what can be its origin? Oxygen vacancies are the principal suspect. By distorting the TiO$_6$ octahedra, they can generate  degeneracies in different degrees of freedom. Moreover, they are suspected to host $d^0$ magnetism \cite{Coey2019}. One cannot forget, however, the unavoidable presence of magnetic impurities at the ppm level in SrTiO$_3$ \cite{Blachly1992,Coey2016}.

The absence of this upturn below $5\times10^{19}$ cm$^{-3}$ and its gradual appearance at higher densities begs explanation. The evolution of the T-square resistivity prefactor may provide one. In presence of a large temperature-dependent resistivity, a small sub-percent upturn may be undetectable. However, this hypothesis fails to explain the gradual fading of the effect  at even higher carrier concentrations. Why the upturn is most prominent when  $n \simeq 2 \times 10 ^{20}$ cm$^{-3}$, which roughly corresponds to one vacancy out of 100 oxygen atoms? A possible clue is provided by the well-established  fact that in general, the Kondo physics of magnetic impurities above a threshold concentration is replaced by a spin-glass state. In the case of iron impurities in silver \cite{Herrera2011} this threshold is about a percent. The same may happen to the Kondo effect of oxygen vacancies. Note that this is yet another argument against extrinsic magnetic impurities as the source of the Kondo effect.

In summary, we found that the existence of the FE order in insulating Sr$_{1-x}$Ca$_x$TiO$_3$ deeply affects charge transport in dilute metallic Sr$_{1-x}$Ca$_x$TiO$_{3-\delta}$. Below a threshold of carrier concentration $n^*(x)$, inelastic and elastic scattering of carriers are both amplified. This threshold density scales with Ca content. Both these features can be explained by invoking the survival of electric dipole interaction in presence of mobile electrons.

Inside the Fermi sea, the dipolar counterpart of RKKY interaction can give rise to a strong anti-ferroelectric coupling at a carrier density proportional to the dipole density. There is striking match between the experimentally resolved proportionality (7.9 dipoles per electron) and what is expected in the simplest RKKY dipolar picture.

At higher carrier concentrations, deep inside the metallic state, clear but tiny upturns of resistivity are detectable, which are reminiscent of the Kondo effect and most probably associated with oxygen vacancies.

\section{METHODS}
SrTiO$_3$ and Sr$_{1-x}$Ca$_x$TiO$_3$ single-crystals with $x= 0.22\%, 0.45\%$, $0.9\%$ were commercially obtained. Mobile electrons were introduced by generating oxygen vacancies. In order to remove oxygen atoms, samples were sealed with high-purity Ti powder ($99.99\%$) in an evacuated quartz tube and annealed in an oven. Samples with various electron carrier concentrations (from $10^{18}$ to $10^{21}$  cm$^{-3}$) were obtained by varying the annealing temperature between 700 and 900 $^\circ$C. Resistivity was measured through the four-terminal method down to 1.8K in Quantum Design PPMS. Hall effect is measured concomitantly under magnetic field to determine the carrier density ($n$).

\section{DATA AVAILABILITY}

The data that support the findings of this study are available from the corresponding author upon reasonable request.

\begin{acknowledgments}
We thank Jonathan Ruhman for stimulating discussions.
This research is supported by National Natural Science Foundation of China via Project 11904294 and Zhejiang Provincial Natural Science Foundation of China under Grant No. LQ19A040005. We acknowledge support by the DFG (German Research Foundation) via Project No. 277146847 within CRC 1238 (subprojects A02, B01, B02, and B03). This work is part of a DFG-ANR project funded by Agence Nationale de la Recherche (ANR-18-CE92-0020-01) and by the DFG through projects LO 818/6-1 and HE 3219/6-1.
\end{acknowledgments}

\section{Competing interests}

The authors declare no Competing Financial or Non-Financial Interests.

\section{AUTHOR CONTRIBUTION}

J.L.W., L.W.Y., C.W.R. and X.L. prepared oxygen-deficient samples. J.L.W., C.W.R. and X.L. performed resistivity measurements. They were all assisted in the measurements by Z.R., T.L. and J.H.. J.L.W., Z.K.X. and X.L. plotted the figures. X.L. and K.B. led the project and wrote the paper. All authors contributed to the discussion.

\section{ADDITIONAL INFORMATION}

Supplementary information accompanies the paper on the npj Quantum Materials website (...).


\begin{thebibliography}{0}%
\makeatletter
\providecommand \@ifxundefined [1]{%
 \@ifx{#1\undefined}
}%
\providecommand \@ifnum [1]{%
 \ifnum #1\expandafter \@firstoftwo
 \else \expandafter \@secondoftwo
 \fi
}%
\providecommand \@ifx [1]{%
 \ifx #1\expandafter \@firstoftwo
 \else \expandafter \@secondoftwo
 \fi
}%
\providecommand \natexlab [1]{#1}%
\providecommand \enquote  [1]{``#1''}%
\providecommand \bibnamefont  [1]{#1}%
\providecommand \bibfnamefont [1]{#1}%
\providecommand \citenamefont [1]{#1}%
\providecommand \href@noop [0]{\@secondoftwo}%
\providecommand \href [0]{\begingroup \@sanitize@url \@href}%
\providecommand \@href[1]{\@@startlink{#1}\@@href}%
\providecommand \@@href[1]{\endgroup#1\@@endlink}%
\providecommand \@sanitize@url [0]{\catcode `\\12\catcode `\$12\catcode
  `\&12\catcode `\#12\catcode `\^12\catcode `\_12\catcode `\%12\relax}%
\providecommand \@@startlink[1]{}%
\providecommand \@@endlink[0]{}%
\providecommand \url  [0]{\begingroup\@sanitize@url \@url }%
\providecommand \@url [1]{\endgroup\@href {#1}{\urlprefix }}%
\providecommand \urlprefix  [0]{URL }%
\providecommand \Eprint [0]{\href }%
\providecommand \doibase [0]{https://doi.org/}%
\providecommand \selectlanguage [0]{\@gobble}%
\providecommand \bibinfo  [0]{\@secondoftwo}%
\providecommand \bibfield  [0]{\@secondoftwo}%
\providecommand \translation [1]{[#1]}%
\providecommand \BibitemOpen [0]{}%
\providecommand \bibitemStop [0]{}%
\providecommand \bibitemNoStop [0]{.\EOS\space}%
\providecommand \EOS [0]{\spacefactor3000\relax}%
\providecommand \BibitemShut  [1]{\csname bibitem#1\endcsname}%
\let\auto@bib@innerbib\@empty
\end{thebibliography}%


\nocite{*}
\begin{thebibliography}{}
 \bibitem{Anderson1965} Anderson, P. W. $\&$ Blount, E. I. Symmetry considerations on martensitic transformations:`` Ferroelectric'' metals? \emph {Phys. Rev. Lett.} \textbf{14}, 217-219 (1965).

 \bibitem{Shi2013} Shi, Y. G. \textit{et al.} A ferroelectric-like structural transition in a metal. \emph{Nat. Mat.} \textbf{12}, 1024-1027 (2013).

 \bibitem{Puggioni2014}
Puggioni, D. $\&$ Rondinelli, J. M. Designing a robustly metallic noncenstrosymmetric ruthenate oxide with large thermopower anisotropy. \emph{Nat. Comm.} \textbf{5}, 3432 (2014).

\bibitem{Laurita2019} Laurita, N. J. \textit{et al.} Evidence for the weakly coupled electron mechanism in an Anderson-Blount polar metal. \emph{Nat. Comm.} \textbf{10}, 3217 (2019).


 \bibitem{Kvyatkovskii2001} Kvyatkovskii, O. E. Quantum effects in incipient and low-temperature ferroelectrics (a review). \emph{Phys. Solid State} \textbf{43}, 1401-1419 (2001).

 \bibitem{Hohnke1972} Hohnke, D. K., Holloway, H. $\&$ Kaiser, S. Phase relations and transformations in the system PbTe-GeTe. \emph{J. Phys. Chem. Solids} \textbf{33}, 2053-2062 (1972).

\bibitem{Alperin1972} Alperin, H. A., Pickart, S. J., Rhyne, J. J. $\&$ Minkiewicz, V. J. Softening of the transverse-optic mode in PbTe. \emph{Phys. Lett. A} \textbf{40}, 295-296 (1972).

\bibitem{Sugai1979} Sugai, S., Murase, K., Tsuchihira, T. $\&$ Kawamura, H. Interaction of the TO-phonon with the acoustic phonons near the phase transition temperature in Pb$_{1-x}$Ge$_x$Te. \emph{J. Phys. Soc. Jpn.} \textbf{47}, 539-546 (1979).

\bibitem{Katayama1987} Katayama, S. I., Maekawa, S. $\&$ Fukuyama, H. Kondo-like effect of atomic motion on resistivity in Pb$_{1-x}$Ge$_x$Te. \emph{J. Phys. Soc. Jpn.} \textbf{56}, 697-705 (1987).

\bibitem{Takaoka1979} Takaoka, S. $\&$ Murase, K. Anomalous resistivity near the ferroelectric phase transition in (Pb, Ge, Sn) Te alloy semiconductors. \emph{Phys. Rev. B} \textbf{20}, 2823-2833 (1979).

\bibitem{Takano1984} Takano, S., Kumashiro, Y. $\&$ Tsuji, K. Resistivity anomalies in Pb$_{1-x}$Ge$_x$Te at low temperatures. \emph{J. Phys. Soc. Jpn.} \textbf{53}, 4309-4314 (1984).

\bibitem{Yaraneri1981} Yaraneri, H., Grassie, A. D. C., Yusheng, H. $\&$ Loram, J. W.  A quasi-Kondo effect in Pb$_{1-x}$Ge$_x$Te alloys. \emph{Phys. C (Solid State Phys.)} \textbf{14}, L411-L444 (1981).

\bibitem{Narayan2019} Narayan, A., Cano, A., Balatsky, A. V. $\&$ Spaldin, N. A. Multiferroic quantum criticality. \emph{Nat. Mat.} \textbf{18}, 223-228 (2019).

\bibitem{Chandra2017} Chandra, P., Lonzarich, G. G., Rowley, S. E. $\&$ Scott, J. F. Prospects and applications near ferroelectric quantum phase transitions: a key issues review. \emph{Rep. Prog. Phys.} \textbf{80}, 112502 (2017).

\bibitem{Muller1979} M\"{u}ller, K. A. $\&$ Burkard, H. SrTiO$_3$: An intrinsic quantum paraelectric below 4 K. \emph{Phys. Rev. B} \textbf{19}, 3593-3602(1979).

\bibitem{Martelli}  Martelli, V., Larrea Jim\'{e}nez, J., Continentino, M., Baggio-Saitovitch, E. $\&$ Behnia, K. Thermal transport and phonon hydrodynamics in strontium titanate. \emph{Phys. Rev. Lett.} \textbf{120}, 125901 (2018)

\bibitem{Bednorz1984} Bednorz, J. G. $\&$ M\"{u}ller, K. A. Sr$_{1-x}$Ca$_x$TiO$_3$: An XY quantum ferroelectric with transition to randomness. \emph{Phys. Rev. Lett.} \textbf{52}, 2289-2292 (1984).

\bibitem{Spinelli2010} Spinelli, A., Torija, M. A., Liu, C., Jan, C. $\&$ Leighton, C.  Electronic transport in doped SrTiO$_3$: Conduction mechanisms and potential applications. \emph{Phys. Rev. B} \textbf{81}, 155110 (2010).

\bibitem{Bhattacharya2016} Bhattacharya, A., Skinner, B., Khalsa, G. $\&$ Suslov, A. V.  Spatially inhomogeneous electron state deep in the extreme quantum limit of strontium titanate. \emph{Nat. Comm.} \textbf{7}, 12974 (2016).

\bibitem{Koonce1967} Koonce, C. S., Cohen, M. L., Schooley, J. F., Hosler, W. R. $\&$ Pfeiffer, E. R. Superconducting transition temperatures of semiconducting SrTiO$_3$. \emph{Phys. Rev.} \textbf{163}, 380-390 (1967).

\bibitem{Lin2013} Lin, X., Zhu, Z., Fauqu\'{e}, B. $\&$ Behnia, K. Fermi surface of the most dilute superconductor. \emph{Phys. Rev. X} \textbf{3}, 021002 (2013).

\bibitem{Collignon2018} Collignon, C., Lin, X., Rischau, C. W., Fauqu\'{e}, B. $\&$ Behnia, K. Metallicity and superconductivity in doped strontium titanate. \emph{Ann. Rev. Cond. Mat. Phys.} \textbf{10}, 25-44 (2019).

\bibitem{Willem2017} Rischau, C. W. \textit{et al.} A ferroelectric quantum phase transition inside the superconducting dome of Sr$_{1-x}$Ca$_x$TiO$_{3-\delta}$. \emph{Nat. Phys.} \textbf{13}, 643-648 (2017).

\bibitem{Stucky2016} Stucky, A. \textit{et al.} Isotope effect in superconducting n-doped SrTiO$_3$. \emph{Sci. Rep.} \textbf{6}, 37582 (2016).

\bibitem{Rowley2018} Rowley, S. E. \textit{et al.} Superconductivity in the vicinity of a ferroelectric quantum phase transition.  Preprint at https://arxiv.gg363.site/abs/1801.08121v2 (2018). 

\bibitem{Herrera2018} Herrera, C., Cerbin, J., Dunnett, K., Balatsky, A. V., $\&$ Sochnikov, I. Strain-engineered interaction of quantum polar and superconducting phases. Preprint at https://arxiv.gg363.site/abs/1808.03739 (2018).

\bibitem{Tomiokal2019} Tomioka, Y., Shirakawa, N., Shibuya, K. $\&$ Inoue, I. H. Enhanced superconductivity close to a non-magnetic quantum critical point in electron-doped strontium titanate. \emph{Nat. Comm.} \textbf{10}, 738 (2019).

\bibitem{Ahadi2019} Ahadi, K. \textit{et al.} Enhancing superconductivity in SrTiO$_3$ films with strain. \emph{Sci. Adv.} \textbf{5}, eaaw0120 (2019).

\bibitem{Kedem2016} Kedem, Y., Zhu, J. X. $\&$ Balatsky, A. V. Unusual superconducting isotope effect in the presence of a quantum criticality. \emph{Phys. Rev. B} \textbf{93}, 184507 (2016).

\bibitem{Wolfle2018} W\"{o}lfle, P. $\&$ Balatsky, A. V. Superconductivity at low density near a ferroelectric quantum critical point: Doped SrTiO$_3$. \emph{Phys. Rev. B} \textbf{98}, 104505 (2018).

\bibitem{Edge2015} Edge, J. M., Kedem, Y., Aschauer, U., Spaldin, N. A. $\&$ Balatsky, A. V. Quantum critical origin of the superconducting dome in SrTiO$_3$. \emph{Phys. Rev. Lett.} \textbf{15}, 247002 (2015).

\bibitem{Kanasugi2019} Kanasugi, S. $\&$ Yanase, Y. Multiorbital Ferroelectric Superconductivity in doped SrTiO$_3$. \emph{Phys. Rev. B}  \textbf{100}, 094504 (2019).

\bibitem{Marel2019} Van der Marel, D., Barantani, F. $\&$ Rischau, C. W. Possible mechanism for superconductivity in doped SrTiO$_3$. \emph{Phys. Rev. Research} \textbf{1}, 013003 (2019).

\bibitem{Kozii2019} Kozii, V., Bi, Z. $\&$ Ruhman, J. Superconductivity near a ferroelectric quantum critical point in ultra low-density Dirac materials. \emph{Phys. Rev. X} \textbf{9}, 031046 (2019).

\bibitem{Glinchuk1992} Glinchuk, M. D. $\&$ Kondakova, I. V. Ruderman${-}$Kittel${-}$like interaction of electric dipoles in systems with carriers. \emph{Phys. Stat. Sol. (b)} \textbf{174}, 193-197 (1992).


\bibitem{Engelmayer2019}  Engelmayer, J. \textit{et al.} Ferroelectric order versus metallicity in Sr$_{1-x}$Ca$_x$TiO$_{3-\delta}$($x=0.009$). \emph{Phys. Rev. B} \textbf{100}, 195121 (2019).


\bibitem{Lin2015} Lin, X., Fauqu\'{e}, B. $\&$ Behnia, K. Scalable T$^2$ resistivity in a small single-component Fermi surface. \emph{Science} \textbf{349}, 945-948 (2015).

\bibitem{Marel2011} Van Der Marel, D., van Mechelen, J. L. M. $\&$ Mazin, I. I.  Common Fermi-liquid origin of T$^2$ resistivity and superconductivity in n-type SrTiO$_3$. \emph{Phys. Rev. B} \textbf{84}, 205111 (2011).

\bibitem{Lin2017} Lin, X. \textit{et al.} Metallicity without quasi-particles in room-temperature strontium titanate. \emph{npj Quant. Mat.}  \textbf{2}, 41 (2017).

\bibitem{Engelmayer2019b}
Engelmayer, J. \textit{et al.} Charge transport in oxygen-deficient EuTiO3: The emerging picture of dilute metallicity in quantum-paraelectric perovskite oxides. \emph{Phys. Rev. Materials} \textbf{3}, 051401(R) (2019).

\bibitem{Maslov2017}
Maslov, D. L. $\&$  Chubukov, A. V. Optical response of correlated electron systems. \emph{Rep. Prog. Phys.} \textbf{80}, 026503 (2017).

\bibitem{Ellmer2001} Ellmer, K. Resistivity of polycrystalline zinc oxide films: current status and physical limit. \emph{J. Phys. D: Appl. Phys.} \textbf{34}, 3097-3108 (2001).

\bibitem{Tahar1998} Bel Hadj Tahar, R., Ban, T., Ohya, Y. $\&$ Takahashi, Y. Tin doped indium oxide thin films: Electrical properties. \emph{Journal of Applied Physics} \textbf{83}, 2631-2645 (1998).

\bibitem{Conwell1950} Conwell, E. $\&$ Weisskopf, V. F.  Theory of impurity scattering in semiconductors. \emph{Phys. Rev.} \textbf{77}, 388-390 (1950).

\bibitem{Tufte1967} Tufte, O. N. $\&$ Chapman, P. W. Electron mobility in semiconducting strontium titanate. \emph{Phys. Rev.} \textbf{155}, 796-802 (1967).


\bibitem{Dingle1955} Dingle, R. B. Scattering of electrons and holes by charged donors and acceptors in semiconductors. \emph{Phil. Mag.} \textbf{46}, 831-840 (1955).

\bibitem{Behnia2015} Behnia, K. On mobility of electrons in a shallow Fermi sea over a rough seafloor. \emph{J. Phys.: Condens. Matter} \textbf{27}, 375501 (2015).

\bibitem{Kleemann2000} Kleemann, W., Dec, J., Wang, Y. G., Lehnen, P. $\&$ Prosandeev, S. A. Phase transitions and relaxor properties of doped quantum paraelectrics. \emph{J. Phys. Chem. Sol.} \textbf{61}, 167-176 (2000).
	
\bibitem{Vugmeister1980} Vugmeister, B. E. $\&$ Glinchuk, M. D. Some features of the cooperative behavior of paraelectric defects in strongly polarizable crystals. \emph{JETP}  \textbf{52}, 482-484 (1980).

\bibitem{Vugmeister1990} Vugmeister, B. E. $\&$ Glinchuk, M. D. Dipole glass and ferroelectricity in random-site electric dipole systems. \emph{Rev. Mod. Phys.} \textbf{62}, 993-1026 (1990).

\bibitem{Samara2003} Samara, G. A. The relaxational properties of compositionally disordered ABO$_3$ perovskites. \emph{J. Phys.: Condens. Matter} \textbf{15}, R367-R411 (2003).

\bibitem{Hemberger1996} Hemberger, J. \textit{et al.}  Quantum paraelectric and induced ferroelectric states in SrTiO$_3$. \emph{J. Phys.: Condens. Matter } \textbf{8}, 4673-4690 (1996).

\bibitem{Wang1998} Wang, Y. G., Kleemann, W., Zhong, W. L. $\&$ Zhang, L.  Impurity-induced phase transition in quantum paraelectrics. \emph{Phys. Rev. B} \textbf{57}, 13343-13346 (1998).

\bibitem{Schremmer} Schremmer H., Kleemann W. $\&$ Rytz D., Phys. Rev. Lett. \textbf{62}, 1896-1899 (1989).

\bibitem{Zhang2002} Zhang, L., Kleemann, W. $\&$ Zhong, W. L.  Relation between phase transition and impurity-polarized clusters in Sr$_{1-\beta}$Ca$_\beta$TiO$_3$. \emph{Phys. Rev. B} \textbf{66}, 104105  (2002).

\bibitem{Landauer} Landauer, R. Spatial carrier density modulation effects in metallic conductivity. \emph{Phys. Rev. B} \textbf{14}, 1474-1479 (1976).

\bibitem{Friedel} Friedel, J. Metallic alloys. \emph{Nuovo Cimento} \textbf{7},  287-311 (1958).

\bibitem{Ruderman1954} Ruderman, M. A. $\&$ Kittel, C. Indirect exchange coupling of nuclear magnetic moments by conduction electrons. \emph{Phys. Rev.} \textbf{96}, 99-102 (1954).

\bibitem{Kondo1964} Kondo, J. Resistance minimum in dilute magnetic alloys. \emph{Prog. Theor. Phys.} \textbf{32}, 37-49 (1964).

\bibitem{Glazman} Glazman, L. I. $\&$ Raikh, M. E. Resonant Kondo transparency of a barrier with quasilocal impurity states. \emph{JETP Lett.}  \textbf{47}, 452-455 (1988).

\bibitem{Fernandez} Fern\'andez-Torrente, I., Franke, K. J. $\&$ Pascual, J. I. Vibrational Kondo effect in pure organic charge-transfer assemblies. \emph{Phys. Rev. Lett.} \textbf{101}, 217203 (2008).

\bibitem{Jarillo} Jarillo-Herrero, P. \textit{et al.} Orbital Kondo effect in carbon nanotubes. \emph{Nature} \textbf{434}, 484-488 (2005).

\bibitem{Matsushita} Matsushita, Y., Bluhm, H., Geballe, T. H. $\&$ Fisher, I. R.  Evidence for charge Kondo effect in superconducting Tl-doped PbTe. \emph{Phys. Rev. Lett.} \textbf{94}, 157002 (2005).

\bibitem{Coey2019} Coey, J. M. D. Magnetism in d$^0$ oxides. \emph{Nat. Mat.} 18, 652-656 (2019).

\bibitem{Blachly1992} Blachly, M. A. $\&$ Giordano, N. Kondo effect in one-dimensional Au (Fe). \emph{Phys. Rev. B} \textbf{46}, 2951-2957 (1992).

\bibitem{Coey2016} Coey, J. M. D., Venkatesan, M. $\&$ Stamenov, P.  Surface magnetism of strontium titanate. \emph{J. Phys.: Condens. Matter} \textbf{28}, 485001 (2016).

\bibitem{Herrera2011} Herrera, W. T. \textit{et al.} Kondo effect and spin-glass behavior of dilute iron clusters in silver films. \emph{Phys. Rev. B} \textbf{84}, 014430 (2011).

\end{thebibliography}
\end{document}